\documentclass[referee]{raa}
\usepackage{graphicx,times}             
\usepackage{natbib}
\usepackage{amssymb,amsmath}
\usepackage{multirow}
\usepackage{booktabs}
\usepackage{color}
\usepackage{ulem}
\bibpunct{(}{)}{;}{a}{}{,}

\usepackage[a4paper=true,dvipdfm=true,pagebackref=true]{hyperref}
\hypersetup{colorlinks = true, linkcolor = green, anchorcolor = red, citecolor = blue, filecolor = red, pagecolor = red, urlcolor = red}

\begin{document}
   \title{Site-testing at Muztagh-ata site II: Seeing statistics
}

   \volnopage{Vol.0 (20xx) No.0, 000--000}      
   \setcounter{page}{1}          

   \author{Jing Xu
      \inst{1,2}
   \and Ali Esamdin
      \inst{1,2}
   \and Jin-xin Hao
      \inst{3}
   \and Jin-min Bai
      \inst{4}
   \and Ji Yang
      \inst{5}
   \and Xu Zhou
      \inst{3}
   \and Yong-qiang Yao
      \inst{3}
   \and Jin-liang Hou
      \inst{6}
   \and Guang-xin Pu
      \inst{1}
   \and Guo-jie Feng
      \inst{1,2}
   \and Chun-hai Bai
      \inst{1}
   \and Peng Wei
      \inst{1}
   \and Shu-guo Ma
      \inst{1}
   \and Abudusaimaitijiang Yisikandee
      \inst{1}
   \and Le-tian Wang
      \inst{1}
   \and Xuan Zhang
      \inst{1}
   \and Liang Ming
      \inst{1}
   \and Lu Ma
      \inst{1}
   \and Jin-zhong Liu
      \inst{1}
   \and Zi-huang Cao
      \inst{2,3}
   \and Yong-heng Zhao
      \inst{3}
   \and Lu Feng
      \inst{3}
   \and Jian-rong Shi
      \inst{3}
   \and Hua-lin Chen
      \inst{7}
   \and Chong Pei
      \inst{7}
   \and Xiao-jun Jiang
      \inst{3}
   \and Jian-feng Wang
      \inst{3}
   \and Jian-feng Tian
      \inst{3}
   \and Yan-jie Xue
      \inst{3}
   \and Jing-yao Hu
      \inst{3}
   \and Yun-ying Jiang
      \inst{3}
}

   \institute{Xinjinag Astronomical Observatory, Chinese Acsdemy of Sciences, Urumqi, 830011, People's Republic of China; {\it xujing@xao.ac.cn,aliyi@xao.ac.cn}\\
        \and  University of Chinese Academy of Sciences,
             Beijing 100049, People's Republic of China\\
        \and
             National Astronomical Observatories, Chinese Academy of Sciences,
             Beijing 100012, People's Republic of China\\
        \and
             Yunnan Observatories, Chinese Academy of Sciences,
             Kunming 650000, People's Republic of China\\
        \and
            Purple Mountain Observatories,Chinese Academy of Sciences,
            Nanjing 210008, People's Republic of China\\
        \and
            Shanghai Astronomical Observatories, Chinese Academy of Sciences,
            Shanghai 200030, People's Republic of China\\
        \and
            National Astronomical Observatories Nanjing Institute of Astronomical Optics \& Technology,
            Chinese Academy of Sciences,
            Nanjing 210008, People's Republic of China\\
\vs\no
   {\small Received~~20xx month day; accepted~~20xx~~month day}}

\abstract{In this article, we present a detailed analysis of the statistical properties of seeing for the Muztagh-ata site which is the candidate site for hosting future Chinese Large Optical/infrared Telescope (LOT) project. The measurement was obtained with Differential Image Motion Monitor (DIMM) from April 2017 to November 2018 at different heights during different periods. The median seeing at 11 meters and 6 meters are very close but different significantly from that on the ground. We mainly analyzed the seeing at 11 meters monthly and hourly, having found that the best season for observing was from late autumn to early winter and seeing tended to improve during the night only in autumn. The analysis of the dependence on temperature inversion, wind speed, direction also was made and the best meteorological conditions for seeing is given.
}

   \authorrunning{J. Xu et al}            
   \titlerunning{Site-testing at Muztagh-ata site II:Seeing statistics }  

   \maketitle


%

\section{Introduction}           
\label{sect:intro}
Muztagh-ata site is one of three potential astronomical locations in western China for hosting future Large Optical/infrared Telescope (LOT) project. LOT, which is an ambitious project with a goal to construct a 12 meters telescope aiming to the frontier scientific research on nature of dark energy, detecting of earth-like extrasolar planets, supermassive black holes, first stars, etc., was elected in 2015\citep{Feng2019}. The site assessment campaign was initiated in January 2017 and lasted for more than two years, climatological properties and optical observing conditions such as sky brightness and cloud amount at Muztagh-ata site have been reported earlier by \cite{Xu2019a} and \cite{Cao2019a}. In this article, we focus on the seeing conditions at our site.

Image quality is directly related to the statistics of the perturbations of the incoming wavefront. With wavefront sensing methods, wavefront fluctuations can be directly analyzed, providing quantitative information on seeing, independent of the telescope being used\citep{1990A&A...227..294S}. Differential image motion monitor (DIMM) has become the standard equipment for assessing the atmospheric “seeing” at astronomical sites \citep{2009PASP..121.1151S}, the seeing derived from DIMM is the combined or integrated effect of all contributing optical turbulence along the optical path\citep{2002PASP..114.1156T}. \cite{2003RMxAC..19...37M} conducted a study with DIMM at San Pedro Mártir observatory and yielded a median seeing of 0.60 $arcsec$ during 123 nights in a three-year period. \cite{2016PASP..128j5003T} measured seeing with a DIMM at Delingha station and achieved an overall seeing median of 1.58 $arcsec$ from 2010 to 2012. Furthermore, DIMM combined with Multi-Aperture Scintillation Sensor (MASS) were widely used for optical turbulence profile measurement \citep{2009PASP..121..922E, 2009PASP..121.1151S, 2012MNRAS.426..635S}.

The seeing measurements was conducted at Muztagh-ata site during the period from April 2017 to November 2018, we analyze the seeing data collected by Differential Image Motion Monitor (DIMM) and give the global, monthly, hourly statistics. We also present the seeing behavior on different conditions of temperature inversion, wind speed and direction. The layout of this paper is as follows: In Section 2, the working method of DIMM and our instruments for seeing measurements are briefly described. In section 3, we introduce how the DIMMs operated at Muztagh-ata site. In Section 4, we firstly show the comparison result between the two DIMMs, then we give the statistics of seeing at different heights during different periods, at last we focus on the 11 meters seeing and detail its statistics from aspects of monthly, hourly trends. Section 5 shows the relation between seeing and some meteorological parameters such as temperature inversion, wind speed and direction. Conclusions are given in Section 6.

\section{Seeing and seeing measurement}
The atmospheric turbulence is usually studied through seeing ($\varepsilon$). The relation between $\varepsilon$, Fried parameter $\gamma_{0}$ , and the turbulence integral is given by \cite{1981PrOpt..19..281R} as:

\begin{center}
\begin{equation}
\label{eq:xj_1}
\varepsilon=0.98\frac{\lambda}{\gamma_{0}}= 5.25 \lambda ^{-1/5}[\int_{0}^{\infty}C_N^2\,dh]^{3/5}
\end{equation}
\end{center}
where $\lambda$ is the wavelength and $\int_{0}^{\infty} C_N^2(h)\,dh$ is the optical turbulence energy profile, $C_N^2$ is refractive index structure constant and $h$ represents altitude.

%
The central wavelength of light measured by the DIMM is $\lambda$ = 0.5 $\mu m$ and the final results have been converted for the direction of observation to zero zenith angle according to the following relation:
\begin{center}
\begin{equation}
\label{eq:xj_2}
\varepsilon_0 = \varepsilon_{0\zeta }\cdot (cos \zeta )^{\frac{3}{5}}
\end{equation}
\end{center}

Where $\zeta$ is the zenith angle, $\varepsilon_{0}$ is the seeing at the zenith, and $\varepsilon_{0\zeta}$ is the seeing as determined by DIMM. The DIMM measurements of the positions of the image centers of stars are made from short exposure images. The exposure time is automatically selected by software according to star magnitude. We focus on the integrated seeing down to the level of the DIMM corrected to one air mass and zero exposure time\citep{2009PASP..121..922E, 2009PASP..121.1151S}.
The Muztagh-ata site was equipped with two DIMMs, one was called French DIMM and the other one was called NIAOT DIMM in the following sections. Parameters of two DIMMs are summarized in Table ~\ref{tab:xj_1}\citep{Wang2019}.

\begin{table*}
\centering
\caption{Parameters and technical specifications of two DIMMs.}
\label{tab:xj_1}
\begin{tabular}{llll}
\hline
              &French DIMM&    &NIAOT DIMM \\
\hline
Telescope Aperture (mm)&300& & 304.8 \\
Focal ratio&8& & 8\\
Focal length (mm)&2400& &1600\\
Sub-apertures&2& &4\\
Sub-aperture diameters (mm)&51& &50\\
Sub-aperture distance (mm)&240& &149\\
Camera&DMK 33GX 174& &Basler aca2040\\
Exposure method&Adjusted automatically& &5ms or 10 ms\\
               &between 0.5 ms and 1000 ms& & \\
Wavelength (mm)&550& &500\\
Output frequency&1 seeing value for& &1 seeing value for every minute\\
                &every 1000 images& &  \\
Scaling and conversion&Convert to zenith&  &Convert to zenith\\
                      &No exposure time scaling& &No exposure time scaling\\
                      &No wavelength scaling& & \\
\hline
\end{tabular}
\end{table*}

\section{Seeing monitoring at Muztagh-ata}
The seeing measurement task at Muztagh-ata site began in April 2017 and lasting until to November 2018. We got median value of 0.83 $arcsec$ from French DIMM and 0.89 $arcsec$ from NIAOT DIMM during the whole measurement period. Here we introduce the two DIMMs' operation during whole measurement period at first.

\begin{figure}
\centering
	\includegraphics[width=12.5cm, height = 4.0cm]{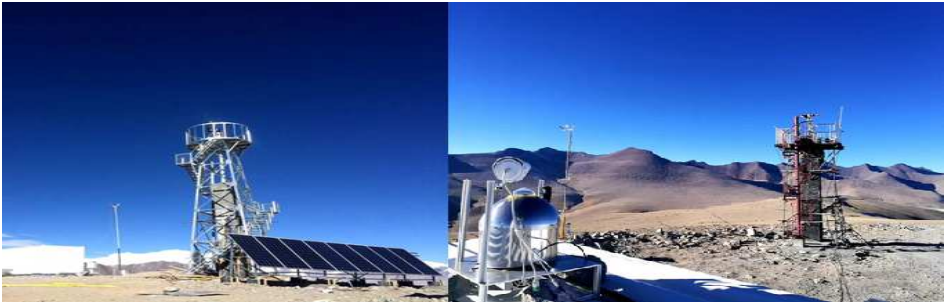}
    \caption{11-meter tower (left) and 6-meter tower (right) at Muztagh-ata.}
    \label{fig:xj_1}
\end{figure}

We have built two towers for seeing measurements, with heights of 11 meters and 6 meters respectively, no dome protected as shown in Figure~\ref{fig:xj_1}. Difference of the heights between the two towers is to explore the effect of difference heights on seeing values. At first the observation was conducted on the ground before two towers were built. Then we moved the two DIMMs to the top of towers successively. The periods of the two DIMMs running at different heights as follow: French DIMM was installed on 15$^{th}$ April 2017, it had been running on the ground until 23$^{rd}$ June 2017, then we moved it to the top of 11-meter tower; NIAOT DIMM was installed on 12$^{th}$ March 2017 and then had been running on the ground until 15$^{th}$ November 2017, then we moved it to the top of 6-meter tower. After the rebuilding of 11-meter tower, for the purpose to ensure two DIMMs can be housed on the top of it simultaneously, we installed NIAOT DIMM on it on 21$^{st}$ September 2018. The two DIMMs were operated at 11 meters until 20$^{th}$ November 2018 for comparing. The detailed operation periods and heights of two DIMMs can be seen in Table~\ref{tab:xj_2}.
\begin{table*}
\centering
	\caption{Operation time periods of the two DIMMs }
	\label{tab:xj_2}
	\begin{tabular}{ccc} 
		\hline
		Starting-Date & French DIMM & NIAOT DIMM \\
		\hline
       2017.03.12 & - & ground \\
       2017.04.15 & ground & ground\\
       2017.06.23 & 11 meters & ground\\
       2017.11.15 & 11 meters & 6 meters\\
       2018.09.21 & 11 meters & 11 meters\\
       \hline
	\end{tabular}
\end{table*}

\begin{figure}
\centering
	\includegraphics[width=12.5cm, angle=0]{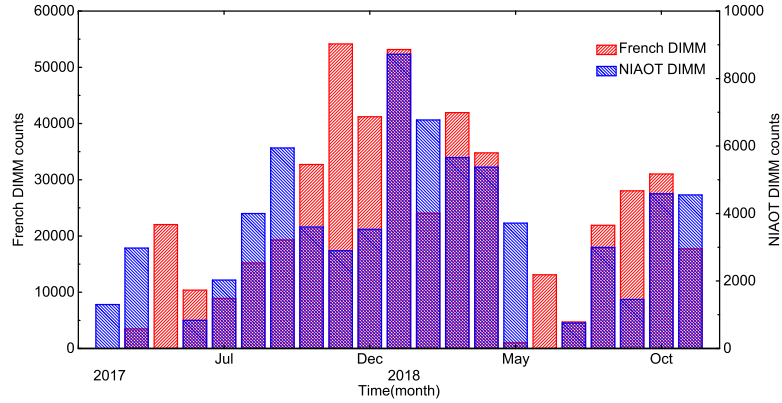}
    \caption{Total monthly data from March 2017 to June 2018. Patterns: French DIMM (red), NIAOT DIMM (blue).}
    \label{fig:xj_2}
\end{figure}

In Figure~\ref{fig:xj_2} we present the total data of the two DIMMs for each month during the acquisition period. Because the sample rate of French DIMM is about five times that of NIAOT DIMM, we use different axes to represent the data amounts of two DIMMs, left for French DIMM and right for NIAOT DIMM.
It is worthy of note that in May 2018 we rebuilt the 11-meter tower so the data amount of French DIMM in that month is relative smaller. Due to the failure of CCD and controller NIAOT DIMM did not operate during May 2017 and June 2018.
\section{Seeing statistics}
\subsection{Comparison of Two DIMMs}
\begin{figure}
\centering
	\includegraphics[width=12.5cm, angle=0]{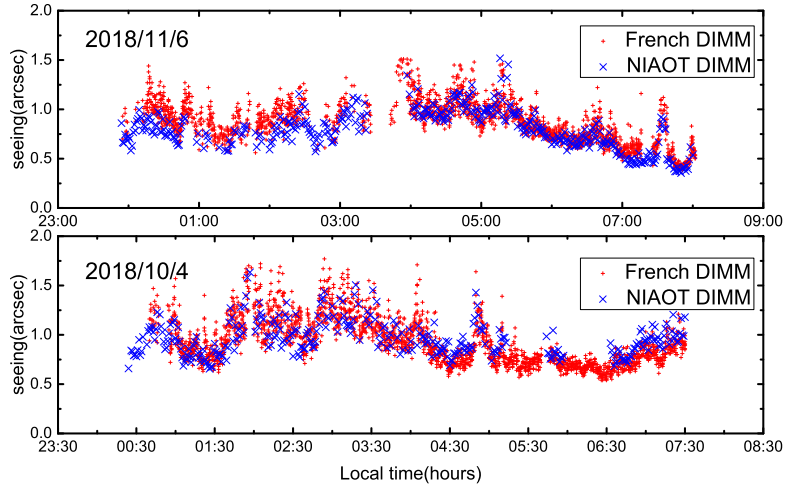}
    \caption{Example of seeing values from French DIMM (red cross) and NIAOT DIMM (blue cross) at 11 meters during two nights.}
    \label{fig:xj_3}
\end{figure}

\begin{figure}
\centering
	\includegraphics[width=12.5cm, angle=0]{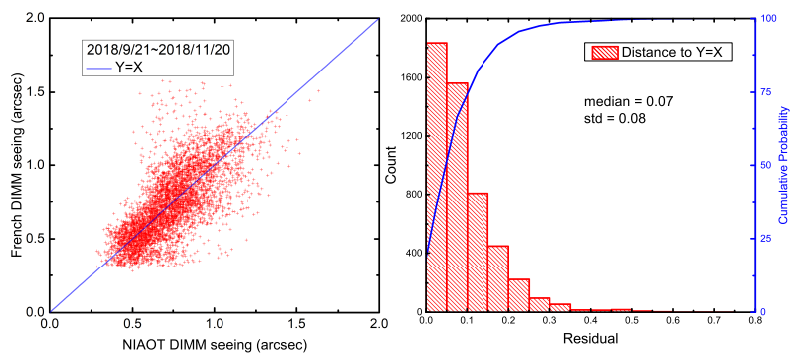}
    \caption{Comparison result of two DIMMs at 11 meters. Left panel: correlation of NIAOT DIMM and French DIMM. Right panel: distribution and cumulative distribution function of residuals.}
    \label{fig:xj_4}
\end{figure}

\begin{figure}
\centering
	\includegraphics[width=12.5cm, angle=0]{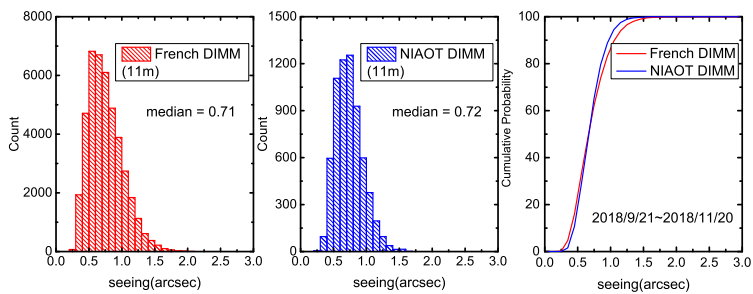}
    \caption{Seeing distributions and cumulative distribution functions from two DIMMs over the period from 21$^{st}$ September 2018 to 20$^{th}$ November 2018 at 11 meters.}
    \label{fig:xj_5}
\end{figure}

Because the 11-meter tower only can accommodate one DIMM before rebuild, so we did the comparison after expanding its platform. The comparison work of French DIMM and NIAOT DIMM was from 21$^{st}$ September 2018 to 20$^{th}$ November 2018, the two DIMMs both were operated on the top of 11-meter tower. In Figure~\ref{fig:xj_3} we show two nights seeing measurement, from which we can see the seeing from French DIMM coincide well with that from NIAOT DIMM. Since the sampling rate of French DIMM is about five times that of NIAOT DIMM, so in this comparison work we firstly find out the nearest moment seeing of French DIMM according to NIAOT DIMM, then use the median value of five-neighborhood as the value of French DIMM. The comparison result of two DIMMs during this period is shown in Figure~\ref{fig:xj_4}, in which the correlation of NIAOT DIMM and French DIMM and the statistical analysis of the residual to line $Y=X$ are given. The median of the residuals is 0.07 and the standard variance is 0.08, it indicates that the difference between the two DIMMs is very small. The seeing distributions and cumulative distribution functions during this period was shown in Figure~\ref{fig:xj_5}, the median seeing values of French DIMM and NIAOT DIMM are 0.71 $arcsec$ and 0.72 $arcsec$ respectively.


\subsection{seeing at different heights}
\begin{figure}
\centering
	\includegraphics[width=12.5cm, angle=0]{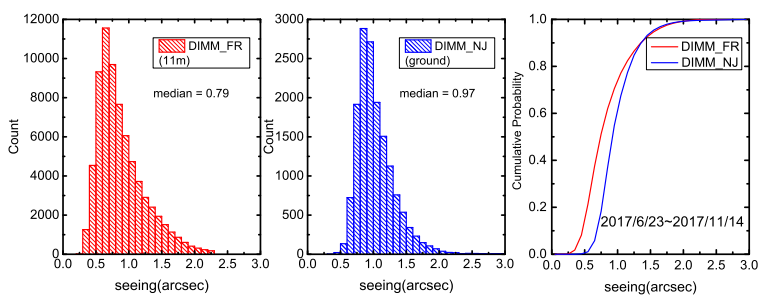}
    \caption{Seeing distributions and cumulative distribution functions from two DIMMs during the period from 23$^{rd}$ June 2017 to 14$^{th}$ November 2017 at 11 meters and on the ground level.}
    \label{fig:xj_6}
\end{figure}

During the whole measurement period we operated two DIMMs at different heights to explore the distribution of near-ground turbulence. In order to eliminate the influence of the occasionally failure of the two DIMMs on results, we only use the data of nights during which both DIMMs were working normally.
From 23$^{rd}$ June 2017 to 14$^{th}$ November 2017, NIAOT DIMM was put on the ground while French DIMM on 11-meter tower. In Figure~\ref{fig:xj_6} we present the distributions and accumulative distribution curves of the seeing values acquired during this period. The median value from French DIMM is 0.79 $arcsec$ at 11 meters and the median value from NIAOT DIMM is 0.97 $arcsec$ on the ground level.

\begin{figure}
\centering
	\includegraphics[width=12.5cm, angle=0]{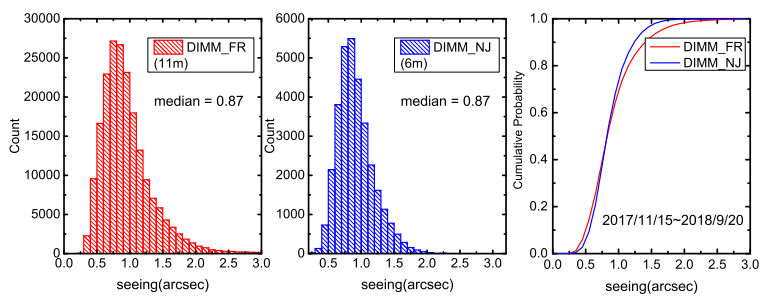}
    \caption{Seeing distributions and cumulative distribution functions from two DIMMs during the period from 15$^{th}$ November 2017 to 20$^{th}$ September 2018 at 11 meters and 6 meters.}
    \label{fig:xj_7}
\end{figure}

In autumn 2017 we built a 6-meter tower, and NIAOT DIMM was moved to the top of 6-meter tower on 15$^{th}$ November 2017. The seeing got from NIAOT DIMM at 6 meters lasted until 20$^{th}$ September 2018. The results of measurement during this period are shown in Figure~\ref{fig:xj_7}. Both of the median seeing values at two heights are 0.87 $arcsec$. The difference between seeing median at 11 meters and ground level is 0.18 $arcsec$, and the difference is very small between 11 meters and 6 meters, indicating that the near-ground turbulence concentrate in the area below 6 meters from ground at Muztagh-ata site.

\subsection{Seeing at 11 meters}
\begin{figure}
\centering
	\includegraphics[width=12.5cm, angle=0]{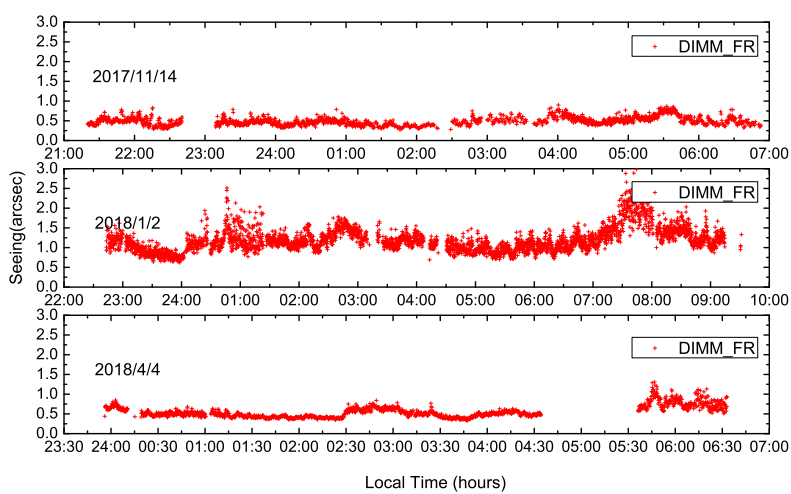}
    \caption{Example of French DIMM seeing values during observable nights. Good steady night (2017-11-14), erratic night (2018-01-02), degrade night (2018-04-04) from top to bottom respectively.}
    \label{fig:xj_8}
\end{figure}

From 23$^{rd}$ June 2017 to 20$^{th}$ November 2018, there were 293 nights in total available for seeing data and 142 nights during which more than 75$\%$ of the time the measurement acquired data.Figure~\ref{fig:xj_8} shows three seeing cases of the 142 nights: good steady nights, with excellent seeing throughout the night (standard variance value less than 0.25 and median value less than 0.82, 66 good steady nights in total, 14$^{th}$ November 2017 for example);erratic seeing night, with irregular seeing throughout the night (standard variance greater than 0.25, 40 erratic nights in total, 2$^{nd}$ January 2018 for example); degrade night in which there is a sudden burst of bad seeing (the standard variance value becomes greater than 0.25 toward dawn, 4$^{th}$ April 2018 for example).

\begin{figure}
\centering
	\includegraphics[width=12.5cm, angle=0]{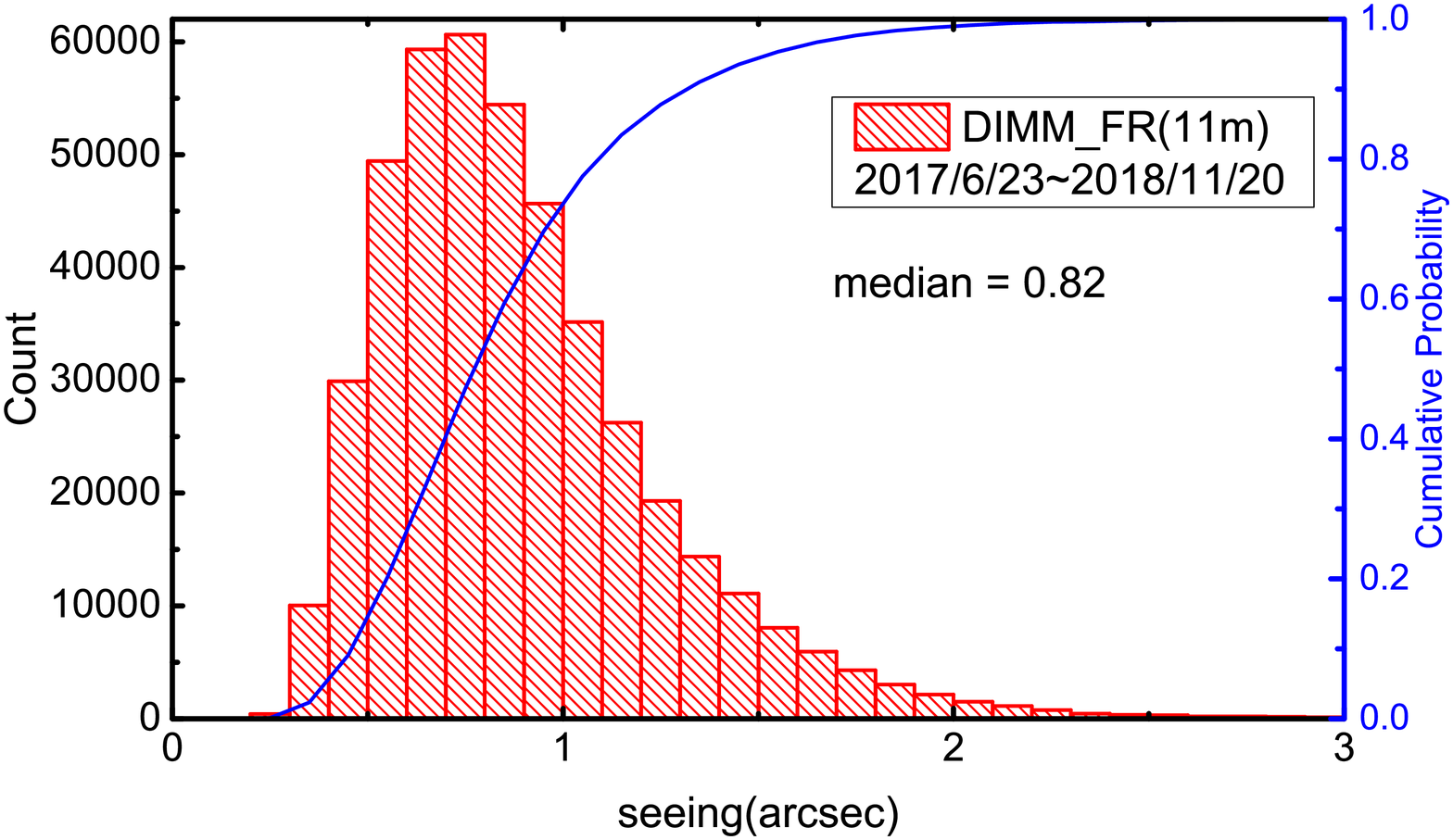}
    \caption{Seeing distributions and cumulative functions from French DIMM at 11 meters during period of 23$^{th}$ June 2017 to 20$^{th}$ November 2018.}
    \label{fig:xj_9}
\end{figure}

\begin{figure}
\centering
	\includegraphics[width=12.5cm, angle=0]{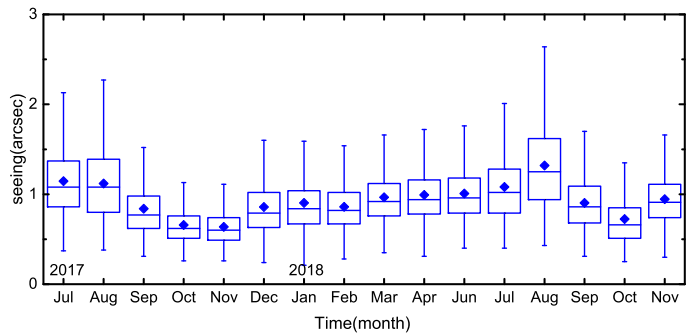}
    \caption{Monthly statistics of seeing from 2017 July to 2018 November (there was no data in 2018 May because of tower rebuilding). Each box represents the values in the range of 25$\%$ to 75$\%$ and the vertical line represents the values ranging from 1$\%$ to 99$\%$, the diamonds and horizontal lines inside every box represent mean and median values respectively.}
    \label{fig:xj_10}
\end{figure}

\begin{figure}
\centering
	\includegraphics[width=12.5cm, angle=0]{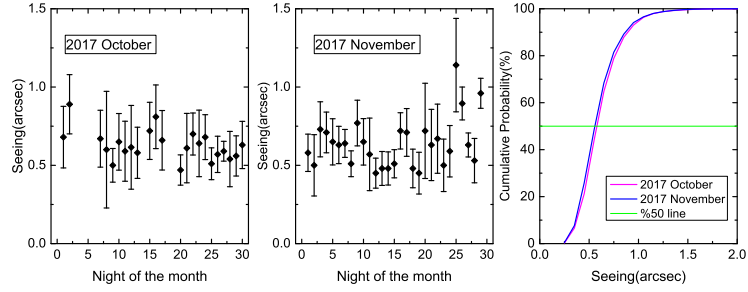}
    \caption{Left and middle panels: Median seeing values (with lower and upper limits represented by first and third quartiles respectively) for each month of October 2017 and November 2017. Right panel: Seeing Cumulative Distribution Function and 50$\%$ line for the two months (Pink for October 2017 and Blue for November 2017).}
    \label{fig:xj_11}
\end{figure}

The French DIMM seeing median value at 11 meters from 23$^{rd}$ June 2017 to 20$^{th}$ November 2018 was 0.82 $arcsec$ as shown in Figure~\ref{fig:xj_9}. We have made a monthly analysis of all available data. There were 16 months from July 2017 to November 2018 available for seeing measurement, no seeing values during May 2018 because the 11-meter tower was in rebuilding. The boxplot in Figure~\ref{fig:xj_10} shows the seeing monthly behavior in the period, each box represents the values in the range of 25$\%$ to 75$\%$ and the vertical line represents the values ranging from 1$\%$ to 99$\%$, the diamonds and horizontal lines inside every box represent mean and median values respectively. From Figure~\ref{fig:xj_10} we can see that the best time period for seeing at Muztagh-ata site is from late autumn to early winter.The seeing nightly statistics and Cumulative Distribution Functions during October and November 2017 are shown in Figure~\ref{fig:xj_11}, the median values are 0.62 $arcsec$ and 0.60 $arcsec$ for these two months. Bad seeing nights appeared frequently in July and August every year, mainly due to the erratic weather during autumn.


\begin{figure}
\centering
	\includegraphics[width=12.5cm, angle=0]{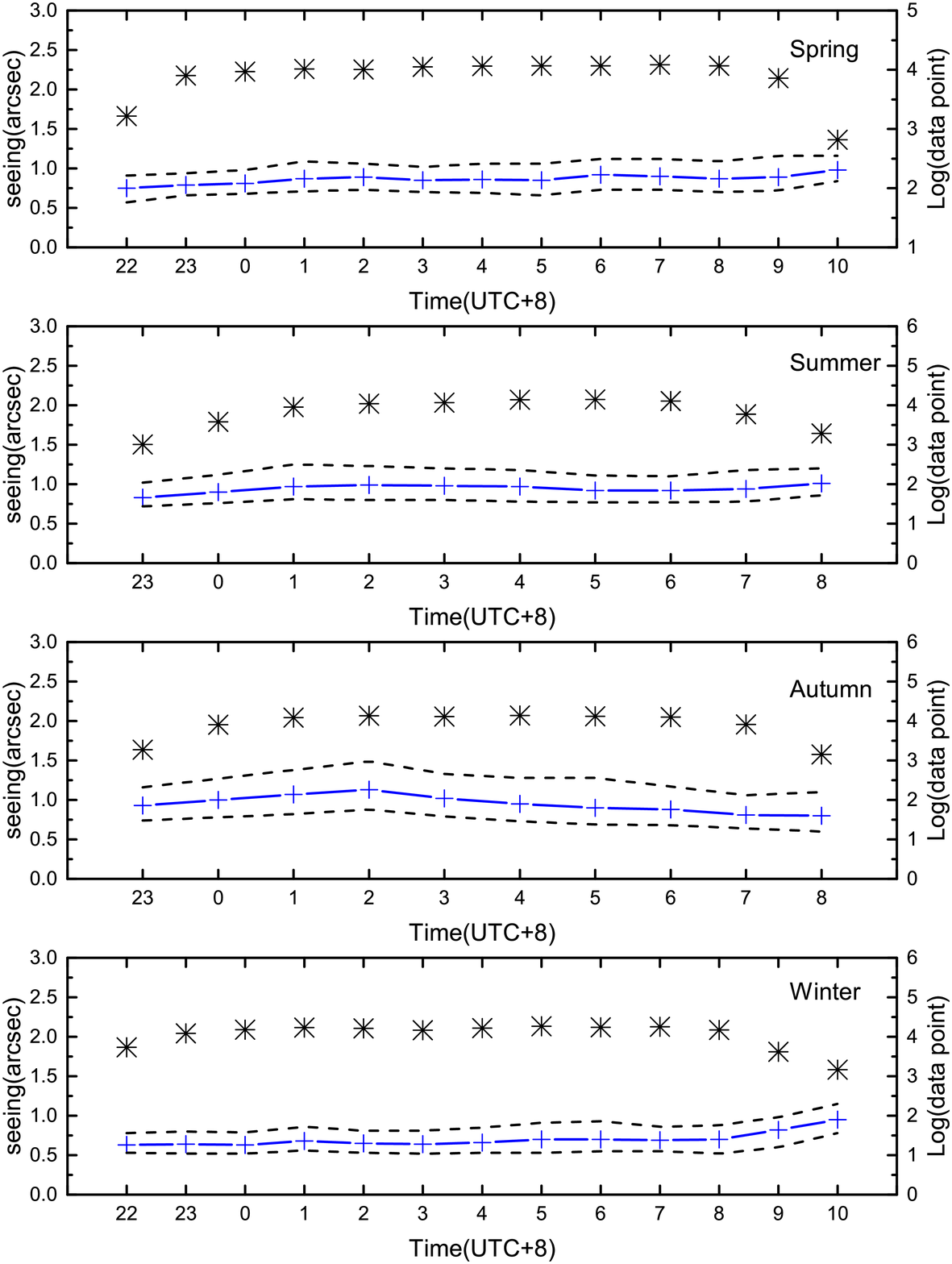}
    \caption{Hourly seeing statistic for each season (spring, summer, autumn, winter from top to bottom respectively). The solid line and pluses indicate the median seeing values in each hour (UT time), and the dashed lines indicate 25$\%$ and 75$\%$. Asterisks mark the amount of data in each hour.}
    \label{fig:xj_12}
\end{figure}

In order to explore the seeing behavior along the night we plot the hourly results for each season integrated over this acquisition period in Figure~\ref{fig:xj_12}(spring, summer, autumn, winter from top to bottom respectively).The solid line and pluses indicate the median seeing values in each hour (UT time), and the dashed lines indicate 25$\%$ and 75$\%$, the asterisks represent the amount of data. In spring, summer and winter the seeing becomes worse toward dawn identically, but in autumn it shows a change that deteriorates first and then improves, it is obviously the seeing during the second half of the night in autumn is better than the first half.

\section{Seeing dependence on meteorological conditions}
One automated weather station named second generation Kunlun Automated Weather Station(KLAWS-2G)\footnote{http://aag.bao.ac.cn/xjtest/index.php} was installed at Muztagh-ata site and started to record data from 1$^{st}$ August 2017. It has several high precision temperature sensors (Young 4-wire RTD ,Model 41342), cup anemometers and wind vanes at different heights make it possible to explore the influence of meteorological parameters on seeing. In this section, the seeing data we use is from French DIMM in the period of 1$^{st}$ August 2017 to 20$^{th}$ November 2018 at 11 meters.
\subsection{Temperature Inversion}
\begin{figure}
\centering
	\includegraphics[width=12.5cm, angle=0]{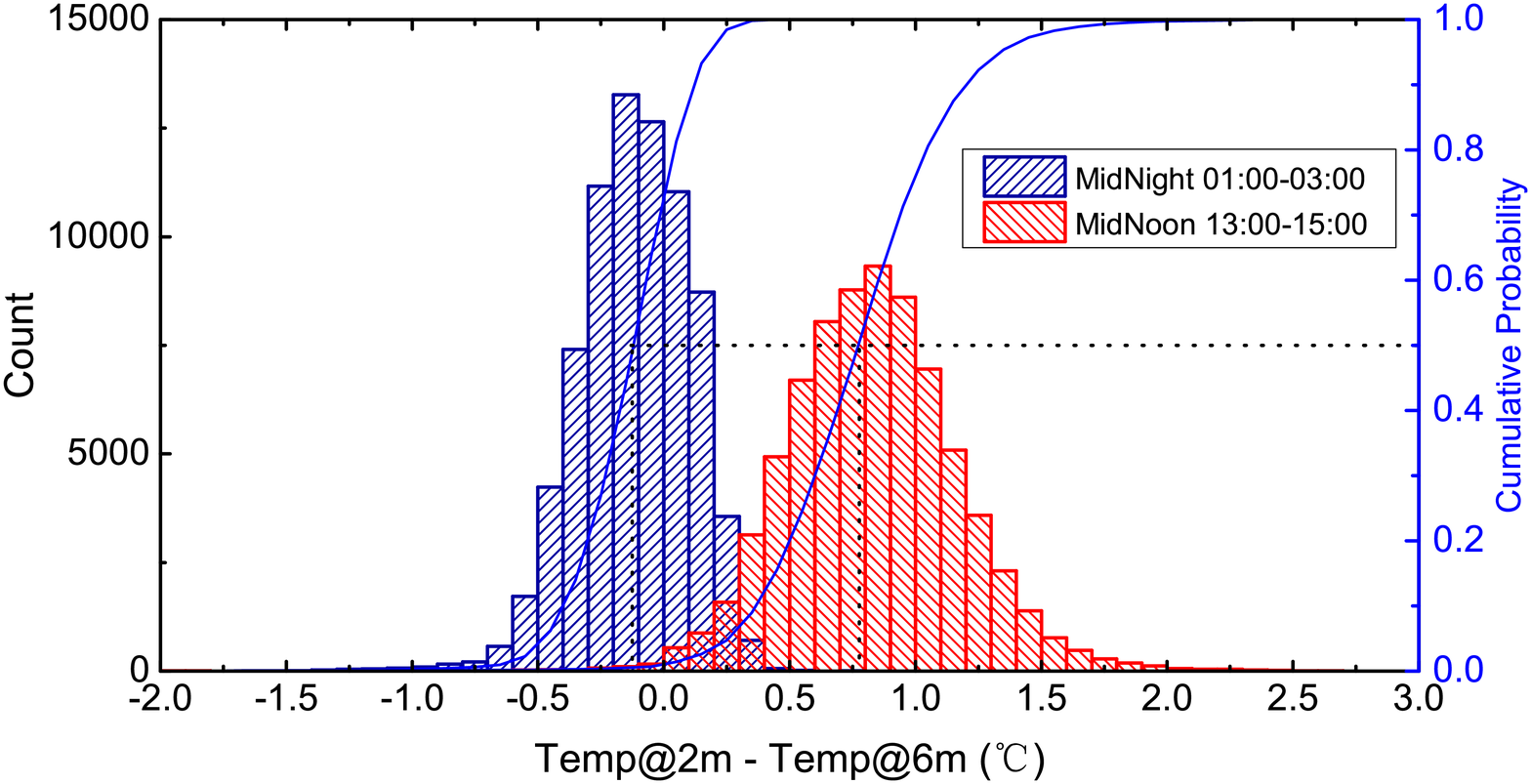}
    \caption{The distribution and cumulative distribution functions of the difference between the temperature at 2 meters and 6 meters in two periods: 13:00~15:00 (UTC+8) afternoon (red) and 1:00$\sim$3:00 (UTC+8) at night (blue).}
    \label{fig:xj_13}
\end{figure}

The phenomenon of temperature inversion means air temperature decreasing with decreasing elevation, it comes with stable atmospheric structure. \cite{2014PASP..126..868H,2019PASP..131a5001H} used KLAWS-2G and found strong temperature inversion occur frequently at Dome A \citep{2010A&ARv..18..417B}. We want to explore the lasting time and influence on seeing at Muztagh-ata.
Figure~\ref{fig:xj_13} shows the statistics of inversion layer in two periods: one is in afternoon from 13:00 to 15:00 (UTC+8), another is in midnight from 1:00 to 3:00 (UTC+8). It can be seen that 70$\%$ of the time during the midnight period the temperatures at 2 meters are smaller than that at 6 meters, and the median value of the difference is -0.2$^{\circ}$$C$.

\begin{figure}
\centering
	\includegraphics[width=12.5cm, angle=0]{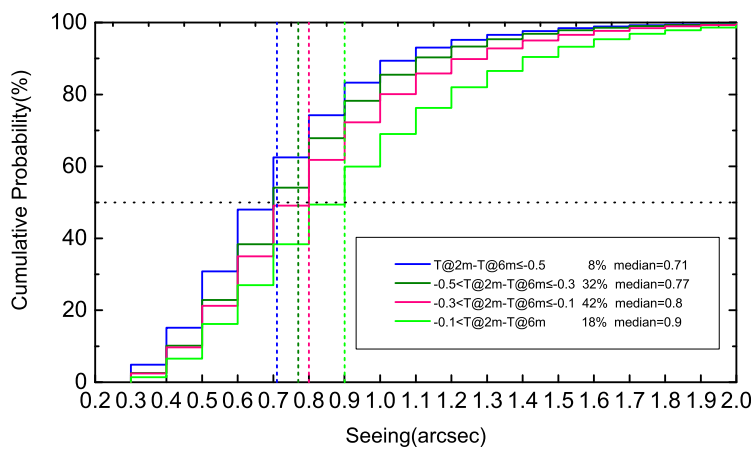}
    \caption{Cumulative distribution functions of seeing in four ranges of the difference between the temperatures at 2 meter and 6 meter: smaller than -0.5 $^{\circ}$$C$ (blue), -0.5 $\sim$-0.3$^{\circ}$$C$ (dark green), -0.3$\sim$-0.1 $^{\circ}$$C$ (pink), higher than -0.1 $^{\circ}$$C$ (light green).}
    \label{fig:xj_14}
\end{figure}

Relationship between seeing and temperature inversion is presented in Figure~\ref{fig:xj_14}, there are four cumulative distribution function curves in the plot represent four ranges of the difference between the temperatures at 2 meters and 6 meters. The blue one is the range of smaller than -0.5 $^{\circ}$$C$ with seeing median value 0.71 $arcsec$. The dark green is the range of -0.5 to -0.3 $^{\circ}$$C$ with seeing median value 0.77 $arcsec$. The pink is the range of -0.3 to -0.1 $^{\circ}$$C$ with seeing median value 0.8 $arcsec$. The light green is the range of higher than -0.1 $^{\circ}$$C$ with seeing median value 0.9 $arcsec$. Figure~\ref{fig:xj_14} proofs that stronger inversion can bring better seeing.

\subsection{Wind}
\begin{figure}
\centering
	\includegraphics[width=12.5cm, angle=0]{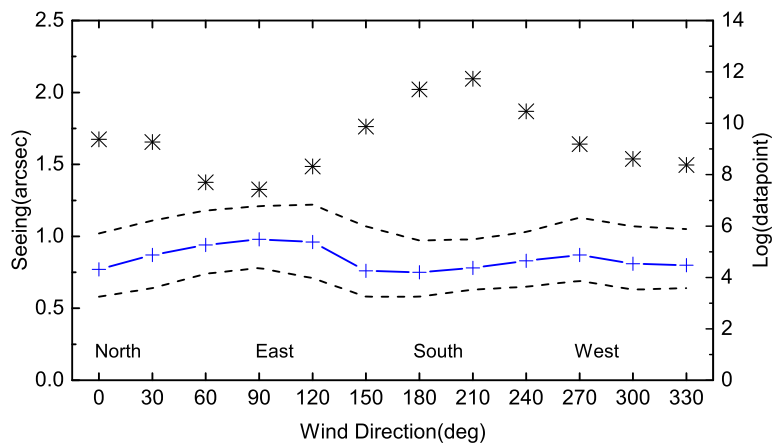}
    \caption{Seeing roses. The solid line and crosses indicate the median seeing values in each 30$^{\circ}$ wind direction bin, the dashed lines indicate 25$\%$ and 75$\%$. Asterisks mark the amount of data in each bin.}
    \label{fig:xj_15}
\end{figure}

\begin{figure}
\centering
	\includegraphics[width=12.5cm, angle=0]{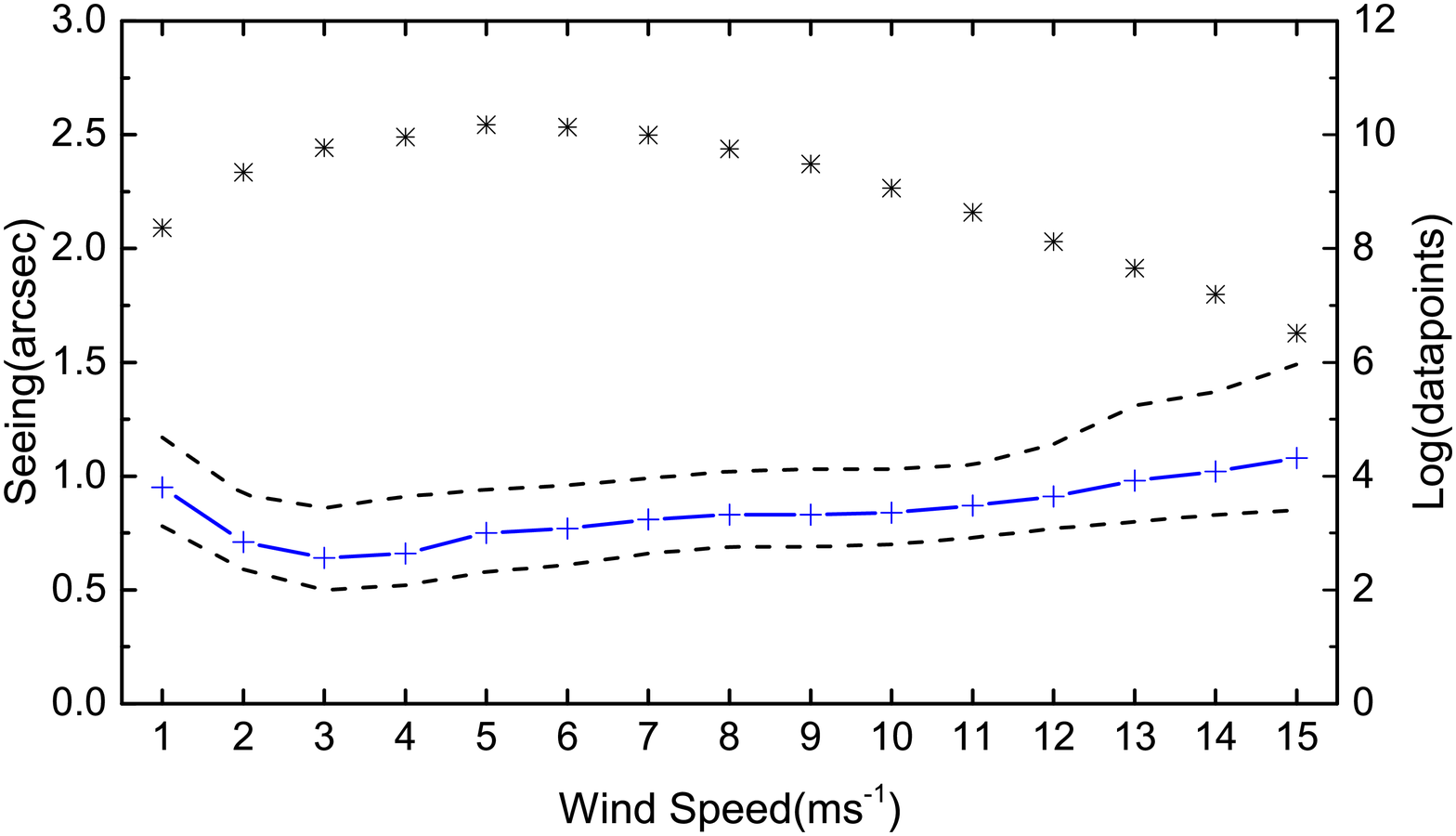}
    \caption{ Seeing statistics at various wind speeds  (wind direction from 180$^{\circ}$ to 270$^{\circ}$). Wind speeds were binned in 1 $ms$$^{-1}$ intervals. The solid line and pluses indicate the median seeing values at various wind speeds, the dashed lines indicate 25$\%$ and 75$\%$. Asterisks mark the amount of data in each bin.}
    \label{fig:xj_16}
\end{figure}

We use the sensors homed at 10 meters altitude for the analysis of the seeing versus wind speed and direction. Figure~\ref{fig:xj_15} and Figure~\ref{fig:xj_16} show the relation of the seeing measurements to wind direction and speed of prevailing wind direction. According to the meteorological parameter measurement results at Muztagh-ata site of this site-testing task, the prevailing wind is southwest in this area\citep{Xu2019a}. The asterisks in Figure~\ref{fig:xj_15} represent the amount of seeing data in every 30$^{\circ}$ of wind direction, the largest amount of data occurs in 180$^{\circ}$ to 240$^{\circ}$ (prevailing wind direction). There are a clear bottom of data amount and a clear peak of median seeing value with east winds. It indicates that the east winds bring about local weather changes and unstable ground turbulence.

Figure~\ref{fig:xj_16} shows the relation of seeing and southwest wind speed (wind direction from 180$^{\circ}$ to 270$^{\circ}$).  The asterisks mark the amount of data and solid line and pluses indicate the median seeing values. With the increase of wind speed, the median seeing value shows an initial drop until the wind speed reaches 3 $ms$$^{-1}$ then keeps growing. When the wind speed is greater than 12 $ms$$^{-1}$ the data amount decreases sharply and 75$\%$ line deviates from median line. From these two figures we can see that the most stable ground turbulence occurs with southwesterly wind speed at 3 $ms$$^{-1}$.

\section{Conclusions}
DIMM data collected from June 2017 to November 2018 at Muztagh-ata site have been presented. We want to acquire some preliminary conclusion of near-ground turbulence distribution and seeing condition at our site. The main results got from this work as follow:

1. The seeing median value at 11-meter during whole measurement period is 0.82 $arcsec$. Seeing difference between 11-meter level and 6-meter level is very small but significant between 6-meter level and ground-level. It illustrates that the near-ground turbulence is concentrated within 6 meters above ground at Muztagh-ata site.

2. Seasonal statistics shows that the best season for optical astronomical observation at Muztagh-ata site is from late autumn to early winter, autumn is relatively bad for seeing because erratic weather. Hourly analysis shows that there is a tendency of seeing that getting worse progressively toward dawn in the most time of year but autumn, in this season the seeing deteriorates first and then improves during night.

3. The dependence of the seeing at 11 meters on meteorological conditions is discussed. We calculate the frequency of temperature inversion during midnight, 70$\%$ of that time the inversion was present. Then we analysis the relationship between inversion and seeing, results present the evidence that stronger inversion can bring better seeing.

4. We present seeing roses and seeing statistics at various wind speeds of prevailing wind direction, from which the conclusion would be made that stable ground turbulence occurs when stable breeze from southwest.

\begin{acknowledgements}
This work is support by the program of the National Nature Science Foundation of China:11873081, 11603065 and the Operation, Maintenance and Upgrading Fund for Astronomical Telescopes and Facility  Instruments, budgeted from the Ministry of Finance of China (MOF) and administrated by the Chinese Academy of Sciences (CAS).
\end{acknowledgements}


\label{lastpage}
\end{document}